\documentclass[pre,showpacs,twocolumn,longbibliography]{revtex4-1}

\usepackage{hyperref}
\usepackage{color}
\usepackage[usenames,dvipsnames]{xcolor}
\usepackage{amsmath,amsthm,amssymb}
\usepackage{graphicx}
\usepackage{epsfig}
\usepackage{dcolumn}% Align table columns on decimal point
\usepackage{bm}% bold math
\usepackage{mathrsfs}
\usepackage{multirow}
\usepackage[all]{xy}
\usepackage{pbox}
\usepackage{verbatim}
\usepackage[latin1]{inputenc}
\usepackage{braket}
\usepackage{mathtools}
\usepackage{bm}
\usepackage{tikz}

\usepackage{mathtools}

\usepackage{mathtools}

\DeclareMathOperator{\Tr}{Tr}

\def\(({\left(}
\def\)){\right)}
\def\[[{\left[}
\def\]]{\right]}

\newcommand{\be}{\begin{equation}}
\newcommand{\ee}{\end{equation}}
\newcommand{\ben}{\begin{eqnarray}}
\newcommand{\een}{\end{eqnarray}}
\newcommand{\beq}{\begin{equation}}
\newcommand{\eeq}{\end{equation}}

\newcommand{\cf}{{\em cf.}\ }
\newcommand{\ie}{{\em i.e.}}

\DeclarePairedDelimiter\floor{\lfloor}{\rfloor}

\newcommand{\xddots}{%
  \raise 4pt \hbox {.}
  \mkern 6mu
  \raise 1pt \hbox {.}
  \mkern 6mu
  \raise -2pt \hbox {.}
}

\begin{document}

\title{Strong zero modes in a class of generalised Ising spin ladders with plaquette interactions}

\author{Loredana M. Vasiloiu}
\author{Federico Carollo}
\author{Matteo Marcuzzi}
\author{Juan P. Garrahan}

\affiliation{School of Physics and Astronomy}
\affiliation{Centre for the Mathematics and Theoretical Physics of Quantum Non-Equilibrium Systems,
University of Nottingham, Nottingham, NG7 2RD, UK}

\date{\today}

\begin{abstract}
We study a class of spin-$1/2$ quantum ladder models with generalised plaquette interactions in the presence of a transverse field. We show that in certain parameter regimes these models have strong zero modes responsible for the long relaxation times of edge spins. By exploiting an infinite set of symmetries in these systems, we show how their Hamiltonians can be represented, in each symmetry sector, by a transverse field Ising chain. Due to the presence of an extensive number of conserved quantities, even if the original system has no disorder, most of these symmetry sectors feature a quasi-random transverse field profile. 
This representation of the ladder system in terms of a disordered Ising chain allows to explain the features of the edge autocorrelation function of the original system. 
\end{abstract}

\maketitle 

\section{Introduction}
Many-body quantum systems can display slow relaxation and long correlation times due to collective dynamical effects. Slow dynamics and non-ergodicity in quantum systems is receiving much attention nowadays both from the experimental and the theoretical side within physics. From a practical perspective, the existence of long time-scales in the dynamics is of interest as it can be related to the possibility of maintaining long coherence times for localised degrees of freedom in many-body systems, with the technological potential for storing and processing quantum information encoded in many-body states \cite{Sarma, Beenakker, JAlicea, Mazza, Bravyi, Ippoliti}. 
On the other hand, the behavior displayed by these systems is connected with topics of fundamental interest in physics such as prethermalization \cite{Else, Essler, Abanin, Kim}, phase transitions \cite{Fendley} and the existence of topological phases of matter \cite{Kitaev, Alicea, Fendley1}. 

Long relaxation time-scales can arise in a variety of physical scenarios. Concrete examples include many-body systems in the presence of disorder leading to many-body localisation \cite{Nandkishore, Huse, Bauer, Bahri, Basko,Znidaric, Pal, Deutsch, Srednicki, Oganesyan, Eisert} and metastability in open quantum systems \cite{Macieszczak,Rose,Gambetta}. 
In this paper we focus on another area where long timescales emerge, quantum systems with free boundaries with so-called strong zero modes (SZMs) \cite{Kemp, Fendley, Fendley1, Sarma, Alicea, Else}. 
Namely, a SZM is an operator which commutes with the dynamical generator up to exponentially small corrections in the system size, the paradigmatic example \cite{Kemp} being the one-dimensional transverse field Ising model (TFIM) with free boundaries \cite{Sachdev, He}: 
in the ferromagnetic phase, one can explicitly construct the SZM operators in an iterative manner and express them as a sum of Majorana fermions \cite{Kemp}. 
The existence of a SZM operator implies the slow relaxation of certain boundary modes, for example giving rise of coherence times that grow exponentially in system size for edge spins in the TFIM \cite{Kemp, Fendley1}. SZMs can also be related to a slow logarithmic-in-time growth of the entanglement entropy \cite{McGinley}. More recently, the possibility of observing SZMs in open settings has also been discussed \cite{Ippoliti,Carmele}, which connects to the more general question of protection of coherences from dissipation \cite{Diehl,Caspel,Buca}. For instance, it can be shown that properly engineered dissipation can increase edge coherence times in spin chains with open boundaries beyond those of the non-dissipative case \cite{JFL}.

\begin{figure}
   \centering
   \includegraphics[width=8.6cm,height=10cm,keepaspectratio]{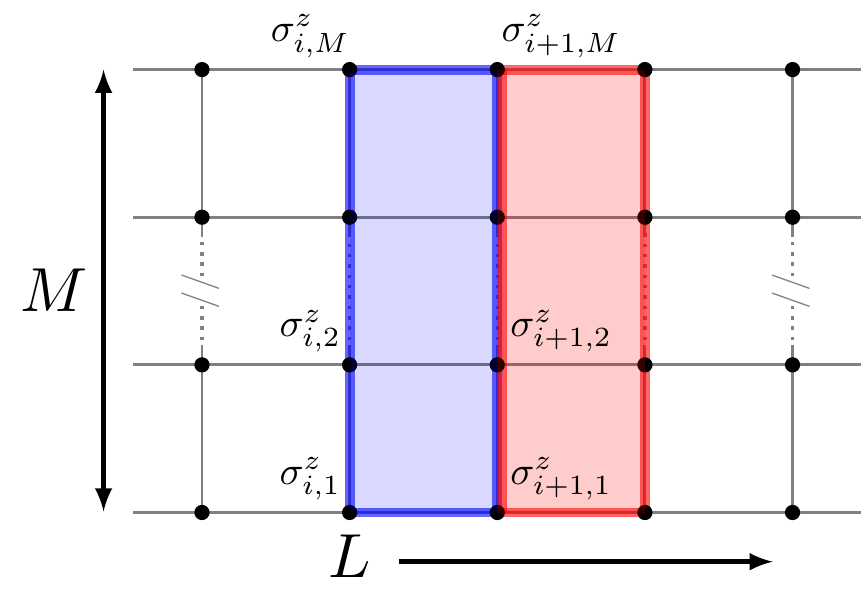}
  \caption{{\bf Quantum $M$-leg ladder model with plaquette interactions}.
  Spins $\sigma^{\mu}_{i,k}$, located at the vertices of an $L\times M$ lattice $(i=1,\dots, L; k=1, \dots, M)$, are subject to a transverse field in the $x$ direction. Interactions (along the $z$ direction) between spins are given by the product of all spin operators $\sigma_{i,k}^{z}$ in two neighbouring rungs of the ladder. Concretely,
  each plaquette is associated with $2M$ spins and the interaction is given by the product $\prod_{k=1}^{M}  \sigma_{i,\,k}^z \, \sigma_{i+1, \,k}^z$. Two such plaquette interactions are shown as the shaded areas in the sketch.}
\label{ladder}
\end{figure}

Here we consider SZMs in a simple generalisation of the TFIM. We consider a class of ladder spin systems with plaquette interactions. Figure~\ref{ladder} describes the class of models we study. The arrangement for an ``$M$-leg ladder'' is that of $M$ parallel chains of length $L$ with a qubit on each site. The interactions are along the $z$ direction involving the product of all sites within one rung of the ladder and the ones on the rung immediately next to it, see the shaded regions in Fig.~\ref{ladder} (details are given below). The system is also subject to a magnetic field on each site in the transverse direction. 

Below we show how to explicitly construct the SZMs in these systems. As we shall see, similarly to what happens in other models \cite{Kemp, Fendley}, the normalisation of the SZMs diverges as the strength of the transverse field approaches a critical value for which the ladder models display a quantum phase transition. 
We then tackle the same class of systems from a different perspective: we identify an extensive set of symmetries of the ladder Hamiltonian which allows us to block-diagonalise them, with each symmetry sector described by a one-dimensional TFIM. Due to the presence of the symmetries, most sectors in general will manifest quasi-randomness in the transverse field. This is another example of how a clean system can mimic the presence of spatial disorder in the Hamiltonian due to a large set of conserved quantities, \cf Refs.~\cite{adamsmith,Smith2018}. We exploit this quasi-random picture to explain the behavior of the autocorrelations of the edge spins in the infinite temperature state.

\section{$M$-leg spin-$1/2$ ladder}

In this section we define the class of models we study. We consider spin systems which consist of a lattice of $M$ parallel chains of $L$ sites each, arranged in such a way that the $i$-th sites of all the chains 
lie on a segment which is perpendicular to the chains themselves, \cf Fig.~\ref{ladder}. Each site of this lattice carries a spin-$1/2$ degree of freedom, and we use the notation $\sigma_{i,k}^{\mu}$ for the local Pauli matrix in the direction $\mu$ acting on the $i$-th site of the $k$-th chain. 

As stated in the introduction, the Hamiltonian of the system is a generalisation of the TFIM Hamiltonian which we define in the following way
\begin{equation}
  H=-J\sum_{i=1}^{L-1} Z_iZ_{i+1} \, 
 -h \sum_{i=1}^L  S_{i}^x ,
 \label{eqladder}
\end{equation}
with 
\begin{equation}
Z_i=\prod_{k=1}^M\sigma_{i,k}^z ,
\end{equation}
and 
\begin{equation}
S_{i}^\mu=\sum_{k=1}^M\sigma_{i,k}^\mu ,
\end{equation}
where the field strength $h$ and coupling strength $J$ are non-negative. The term proportional to $h$ represents the uniform transverse magnetic field, while the terms in the first sum appearing in Eq.~\eqref{eqladder} 
represent longitudinal magnetic plaquette interactions between neighboring rungs of the ladder -- as illustrated in Fig.~\ref{ladder}. Notice that the TFIM is recovered when $M=1$. 

For the whole $M$-leg ladder model we can define a parity operator 
\begin{equation}
 \mathcal{F}= \prod_{i=1}^L \prod_{k=1}^M \sigma_{i,\, k}^x ,
 \label{eqsymmetry}
\end{equation}
which implements a discrete $\mathbb{Z}_2$ spin-flip symmetry. When $M>1$, however, one can also decompose such a parity operator into the product of single chain parity operators 
\begin{equation}
\mathcal{F}_{k}=\prod_{i=1}^L  \sigma_{i,\, k}^x ,
\end{equation}
with $\mathcal{F}=\prod_k \mathcal{F}_k$. 

One can easily check that  
\begin{equation}
\left[ H, \mathcal{F}_k \right]=0 \;\;\; \forall k ,
\end{equation}
so that 
\begin{equation}
\left[ H, \mathcal{F} \right]=0 .
\end{equation}
This implies that an eigenstate $|\varphi\rangle$ of $H$ can be chosen to be a simultaneous eigenvector of all the $\mathcal{F}_k$,
$$
\mathcal{F}_k|\varphi\rangle=\beta_k |\varphi\rangle\, ,  \mbox{with} \, \beta_k=\pm 1\, ,
$$
and we can decompose the Hilbert space into parity sectors $\mathbb{P}_{\vec{\beta}}$, with $\vec{\beta}=(\beta_1,\beta_2,\dots,\beta_M)$, which are mapped onto themselves by the action of the system Hamiltonian.

\section{Strong zero modes in $M$-leg ladders}

After having introduced the class of models we consider, we will now proceed to show that they feature, in certain parameter ranges, SZMs. We first review the definition of a SZM \cite{Kemp, Fendley, Fendley1}, and then explicitly construct the SZMs for the $M$-leg ladders. 

A SZM $\Psi$ is an operator which almost commutes with the Hamiltonian, $\|[H,\Psi]\|\approx \epsilon_L$, with a correction term $\epsilon_L$ which is exponentially decaying with the size $L$ of the system. 
Usually, such an operator is also required to anticommute with a parity operator $P$, commuting with the Hamiltonian $[P,H]=0$. When this is the case, the SZM is an operator that maps an eigenstate of the Hamiltonian in one of the two parity sectors into an eigenstate in the other sector. The two mapped states are split in energy by $\epsilon_L$, and since this mapping occurs across the energy spectrum this guarantees the existence of a long coherence times. 

For general $M$-leg ladder models, however, due to the presence of a larger number of parity sectors, $\mathbb{P}_{\vec{\beta}}$, the situation is more subtle. Indeed, as we show below, SZM operators for this class of systems obey
\begin{equation}
\{\Psi,\mathcal{F}_k\}=0\, , \, \forall \, k \, .
\label{apk}
\end{equation}
In terms of the parity operator for the whole system, $\mathcal{F}$, one then has 
\begin{equation}
\left\{
\begin{array}{rcl} 
[\Psi, \mathcal{F}] = 0 & & M \, {\rm even} \\
\\
\{\Psi,\mathcal{F}\} = 0 & & M \, {\rm odd} \\
\end{array}
\right.
\end{equation}
While the commutation/anticommutation relations of the SZM with $\mathcal{F}$ are not of help to understand the action of the SZM on eigenstates of the Hamiltonian, 
the properties expressed in Eq.~\eqref{apk} can be used to show that it connects eigenstates and eigenvalues of the Hamiltonian in the parity sector $\mathbb{P}_{\vec{\beta}}$ with those of $\mathbb{P}_{-\vec{\beta}}$. To show this, lets assume $|\varphi\rangle$ to be an eigenvector of the Hamiltonian associated to an eigenvalue $\lambda_\varphi$ belonging to the parity sector $\mathbb{P}_{\vec{\beta}}$. One then  has that the vector $\Psi|\varphi\rangle$ is an eigenvector of the Hamiltonian:
\begin{equation}
H\Psi|\varphi\rangle = \lambda_\varphi \Psi|\varphi\rangle+O(\epsilon_L) \, ,
\end{equation}
where we have used the fact that the SZM commutes with the Hamiltonian up to terms of order $\epsilon_L$ exponentially small in system size. In addition, exploiting the anticommutation relations of the SZM with the parities $\mathcal{F}_k$ one has that
$$
\mathcal{F}_k\Psi\ket{\varphi}=-\beta_k\Psi\ket{\varphi}\, , \forall k\, ,
$$
showing indeed that $\Psi\ket{\varphi}\in \mathbb{P}_{-\vec{\beta}}$. 
Summarizing, this means that for large system sizes the spectrum of the Hamiltonian $H$ gets paired up between corresponding parity sectors $\mathbb{P}_{\vec{\beta}}\leftrightarrow\mathbb{P}_{-\vec{\beta}}$. For explicit examples we refer to Appendix \ref{pairing}. 

After having discussed the consequences of the existence of the SZM on the spectral properties of the ladder Hamiltonian we now provide its explicit form, from which we can recover the properties described in the previous discussion. 
As in the case of the TFIM \cite{Kemp}, for general $M$-leg ladder models the SZM can be written as a sum of operators 
\begin{equation}
\Psi= \sum_{j=1}^L  \psi_j\, ,
\label{SZM}
\end{equation}
where the $\psi_j$ are given by 
\begin{equation}
\psi_j=\left(\frac{h}{J} \right)^{j-1} \prod_{i=1}^{j-1} S_i^x  \, Z_j \, .
\label{SZMJ}
\end{equation}
Notice that the terms $\psi_j$ appearing in the SZM are almost Majorana operators for $M \neq 1$: while they anticommute, $\{\psi_j,\psi_i\}=0$ for $i\neq j$, for $i=j$ the anticommutator is not a multiple of the identity (yet commutes with any other term $\psi_k$). 

The first and most important property to be shown is that this operator $\Psi$ almost commutes with the Hamiltonian. 
To do this we note that the Hamiltonian can be split in two parts: one describing the magnetic interaction between spins around plaquettes, $H_0\propto\sum_j Z_jZ_{j+1}$, 
and the other describing the effects of the transverse magnetic field $H_1\propto \sum_j S_j^x$. Thus, we can write the commutator of the Hamiltonian with the SZM in the following way
\begin{equation}
\begin{split}
 \left[ H, \Psi \right] &= \sum_{j=1}^L \left( \left[ H_0, \psi_j \right] + \left[ H_1, \psi_j \right]\right) \\
 &= \sum_{j=1}^{L-1} \left(  A_{j+1} + B_j \right) + B_L,
\end{split}
\label{TelSum}
\end{equation}
where we introduced the quantities $\left[ H_0, \Psi_j \right] =A_j$ and $\left[ H_1, \Psi_j \right] =B_j$, and used $\left[ H_0, \Psi_1 \right]=A_1=0 $.
We show in Appendix \ref{compactform} that this is a telescoping series, namely $A_{j+1} + B_j=0, \forall j$, therefore the only contribution to the commutator comes from the term $B_L$ with norm $\|B_L\| \approx 2J {\alpha}^L$, where $\alpha= M h/J$. This thus proves that the operator defined in Eq.~\eqref{SZM} almost commutes with the Hamiltonian up to a factor that scales as $\alpha^L$. When $\alpha<1$ the correction to the commutator becomes exponentially small with system size. 

The almost commutation of the SZM with the ladder Hamiltonian can be used to show the slow relaxation of coherences for the edge spins. Indeed, when $\alpha<1$, one can notice from Eq.~\eqref{SZMJ} that the SZM is highly localised around the first rung of the ladder. 
In particular, when $h\ll J$, one has that the SZM is $\Psi \approx Z_1$ up to corrections of order $h$: thus exploiting the fact that this SZM almost commutes with the Hamiltonian one can show (in the same way as it happens for the TFIM \cite{Kemp}) that the infinite temperature autocorrelation function 
\begin{equation}
C_{\infty}(t):=\frac{1}{2^{M\,L}}\Tr\left(Z_1(t)Z_1\right)\approx 1 \, ,
\label{ACF}
\end{equation}
up to times of the order $\tau\approx \alpha^{-L}$. We will discuss the behaviour of the autocorrelation in detail below.

\begin{figure}
\centering
\includegraphics[scale=0.855]{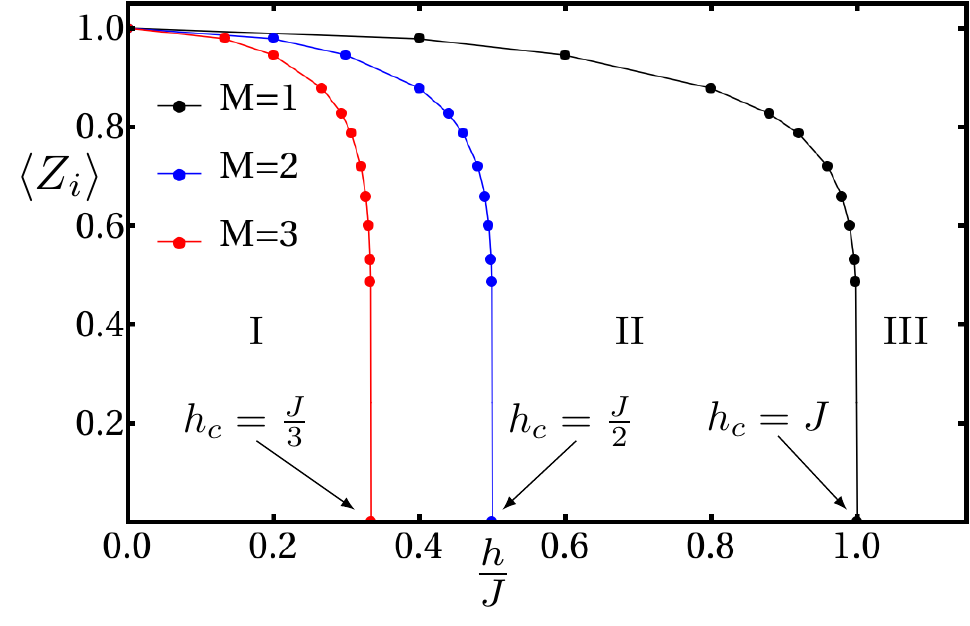}
\caption{{\bf Phase transition in the ladder model for different values of $M$}. The order parameter is the expectation value in the ground state of the product of the magnetization operators along the $z$ direction for the single rungs of the ladder,  $\langle Z_i\rangle$. When $M=1$ (TFIM case) the transition occurs at $h_c=J$. For general values of $M$ the transition is shifted to $h_c=J/M$. We show here results for $M=1,2,3$ obtained by means of infinite time-evolving block decimation algorithms \cite{iTEBD1,iTEBD3,iTEBD4,iTEBD2}. }
\label{Fig2}
\end{figure}

\section{Norm of $\Psi$ and relation to phase transitions}

An important property for the SZM is that it must be normalisable. 
Indeed, if this were not true it would not be possible to give a meaning to the manipulations described in the previous section. 
In this section we prove the conditions under which the SZM operator has finite norm. 
We will also show that the point at which the norm of the SZM becomes unbounded, the ground state of the system undergoes an Ising-type phase transition. This seems to provide further evidence of the relation between existence of SZM and the occurrence of phase transitions in quantum Hamiltonian systems \cite{Fendley}. 

To compute the norm of the SZM, \cf Eq.~\eqref{SZM}, we make use of the anticommutation relations, $\{ \psi_i, \psi_j\}=0$, $\forall i \ne j$, and we express the square of the SZM as 
$\Psi^2=\sum_{i=1}^L \psi^2_i$, with 
$$
\psi^2_i=\left(\frac{h}{J} \right)^{2\left(i-1 \right)} \prod_{k=1}^{i-1} (S_k^x)^2\, .
$$
Since all terms in $\Psi^2$ commute among themselves we can compute the norm as 
\begin{equation}
 \|\Psi\|=\sqrt{\| \Psi^2 \|}=\left( \frac{1-{\alpha}^{2L}}{1-{\alpha}^{2}}\right)^{1/2}\, .
\label{norm}
\end{equation}
From the above equation it is clear that the SZM is normalisable for $\alpha < 1$. On the other hand, one can show that the ladder model undergoes a quantum phase transition when the transverse field approaches the critical value $h_c= J/M$, corresponding to $\alpha=1$. 

The order parameter for an $M$-leg ladder system is the expectation value in the ground state of the product of operators along the $z$ direction for a rung of the ladder, $\langle Z_i\rangle$ (which generalises the order parameter of the TFIM). In Fig.~\ref{Fig2} we plot the order parameter as a function of magnetic field for different values of $M$: we see that the point at which the eigenstates pairing due to the SZM disappears is the same value of the transverse field for which the quantum phase transition in the ground state of the Hamiltonian takes place \cite{Fendley}.

\section{Quasi-random Ising chains from the ladder model}
In this section we show how the Hamiltonian for the ladder model can be decomposed into a direct sum of Ising Hamiltonians with a transverse field that can mimic the presence of randomness in the chain. 
In order to achieve this, in the spirit of \cite{adamsmith}, we characterize an extensive set of conserved quantities which we can then use to decompose the Hamiltonian in the various blocks identified by the conserved charges. 

The first step thus consists in finding the above mentioned symmetries of the Hamiltonian. This can be achieved by directly inspecting the operator $S_i^2= (S_i^x)^2 + (S_i^y)^2 + (S_i^z)^2$, 
proportional to the total angular momentum in each rung of the ladder, and also the square of the magnetization along the $x$ direction $(S_i^x)^2$. One can check that 
\begin{equation}
\begin{split}
  \left[S_i^2, H \right]=0\, , \quad  \forall i\, , \\
 \left[(S_i^x)^2, H \right]=0\, , \quad  \forall i\, .
\end{split}
\end{equation}
The existence of these conserved quantities allows us to represent the original ladder Hamiltonian in a simultaneous eigenbasis of $S_i^2$ and $\left( S_i^x \right)^2$ in the various symmetry sectors identified by the quantum numbers of $S_i^2$ and $(S_i^x)^2$.

We thus need to identify these values and to understand the action of the Hamiltonian on their eigenstates. 
From the definition of $S^{x}_i$ and $S_i^2$ it is clear that we are dealing with the coupling of $M$ spin-$1/2$ systems along each vertical column characterized by the index $i$. 
Therefore, the possible eigenvalues $s_i^2$ of $S_i^2$ are  
\begin{equation}
 s_i^2= 4 \ell(\ell+1),
\end{equation}
where the factor $4$ comes from the fact that we have defined $S_i^2$ in terms of Pauli matrices and not in terms of spin operators. Moreover $\ell_{\rm min} \le \ell \le  \ell_{\rm max} $, where $\ell_{\rm max}=M/2$, while $\ell_{\rm min}=0$ if $M$ is even and $\ell_{\rm min}=1/2$ if $M$ is odd. 
For a fixed irreducible representation of the total angular momentum where $S_i^2=4\ell (\ell +1)$ one can also recover the possible quantum numbers of $(S_i^x)^2$ noticing that the eigenvalues of the magnetization along the $x$ direction, $s_i^x$, belong to the set $[-2\ell,-2\ell+2,\dots 2\ell-2, 2\ell]$. 

Thus, in the sector identified by the choice $\left( s_i^2,(s_i^x)^2 \right)$ for the conserved quantum numbers, the $i$th rung of the ladder can be represented by means of a two-level system with states $|+\rangle$, $|-\rangle$ defined such that 
\begin{equation}
S^2_i|\pm\rangle =s_i^2|\pm\rangle \, ,\qquad S^x_i|\pm\rangle =\pm s_i^x|\pm\rangle\, ,
\end{equation}
if $s_i^x\neq0$. In turn, when $s_i^x=0$ (only possible for $M$ even) the dimension of the irreducible representation of the angular momentum is one and  the rung is described by a single state $|0\rangle$, such that $S_i^x |0\rangle=0$.

In order to complete the mapping from the ladder to the chain, we need to understand the action of $Z_i$ in the reduced space for the rung $i$ identified by a choice of the conserved numbers. 
From the fact that $Z_i$ anticommutes with $S_i^x$, one easily gets 
$$
Z_i|\pm\rangle =|\mp\rangle\, ,
$$
while, if $s_i^x=0$ then $Z_i|0\rangle =\pm|0\rangle$, where the sign depends on the symmetric/antisymmetric property of the singlet $|0\rangle$. 
These considerations thus point to the fact that for a choice of the conserved quantum numbers $\left( s_i^2,(s_i^x)^2 \right)$ the single rung operators are mapped onto 
\begin{equation}
\begin{split}
 S_i^x =s_i^x \, \tau_i^x,\,\,  Z_i=\tau_i^z \qquad &\mbox{ if $s_i^x\neq0$ }; \\
 S _i^x =0,\,\,  Z_i=\pm1 \qquad &\mbox{ if $s_i^x = 0$ } ,
\end{split}
\label{rungmapping}
\end{equation}
where $\tau_i^\mu$ are auxiliary Pauli matrices. 

The last thing that one needs to clarify in order to show how the ladder Hamiltonian can be reduced to an Ising model one with quasirandom potential,  is the multiplicity of the different values of the Hamiltonian quantum numbers. 
Such an information can be recovered from the Clebsch$-$Gordan decomposition series \cite{Zachos}. Indeed, we need to iteratively compose $M$ spin-$1/2$ systems and to consider all the irreducible representations arising from this procedure; 
the above mentioned series takes into account the fact that, each time a spin-$1/2$ system is added to an irreducible representation, we obtain two irreducible representations,
one with dimension increased by one and the other with dimension decreased by one with respect to the original representation.

The Clebsch$-$Gordan series is the following 
\begin{equation*}
\pmb2^{\otimes M}= \bigoplus \limits_{k=0}^{\floor*{\frac{M}{2}}} \left( \frac{M+1-2k}{M+1} \binom{M+1}{k} \right) \left( \pmb{ M +1 -2k} \right),
\end{equation*}
where $\floor*{\frac{M}{2}}$ is the integer floor function and $\binom{x}{y}$ is the binomial coefficient; 
it shows the dimension of the irreducible representations in the Clebsch$-$Gordan reduction of $S_i^2$ (boldface factor on the right side of the above equation) together with their multiplicity given by the first factor in the direct sum. 
 
The dimension of the representation and the value $s_i^2=4\ell(\ell+1)$ are connected through the relation $\ell=(M-2k)/2$; therefore we can say that the eigenvalue $s_i^2=4\ell(\ell+1)$ of $S_i^2$ appears with a multiplicity $\kappa_\ell$
$$
\kappa_\ell= \frac{2\ell+1}{M+1} \binom{M+1}{\frac{M-2\ell}{2}} \, .
$$

We can also recover all possible values of $|s_i^x|$: these are in the form $|s_i^x|=M-2k$, with $0\le k\le \floor*{\frac{M}{2}}$ and can appear with a multiplicity $\nu_{M-2k}$ 
that can be derived from the multiplicity of the irreducible representations in the following way
\begin{equation}
 \nu_{M-2k}= \sum_{m=0}^{k}\frac{M+1-2m}{M+1}\binom{M+1}{m}\, ;
 \label{multiplicity}
\end{equation}
namely, to find how many times the value $M-2k$ can appear for $|s_i^x|$, we need to count all possible irreducible representations containing that specific value. An explicit example is provided in the Appendix \ref{exampleseries}.

In the next subsections we use these findings to map the Hamiltonian of the system into an Ising-like Hamiltonian for a generic symmetry sector identified by fixing the quantum numbers $s_i^2$ and $(s_i^x)^2$ for all rungs in the ladder. The two cases, $M$ odd and $M$ even, present a substantial difference so that we will treat them separately.

\subsection{Block decomposition of the ladder Hamiltonian for $M$ odd}

When the number of spins on each rung $M$ is odd, for any choice of the conserved quantities $S_i^2$ and $(S_i^x)^2$, columns in the ladder behave like two-level systems, \cf Eq.~\eqref{rungmapping}. Therefore, by fixing for each rung $i$ a choice of the values $(s_i^2, (s_i^x)^2)$, the ladder Hamiltonian can be mapped onto an Ising-like Hamiltonian 
\begin{equation}
 \tilde{H}=-J \sum_{i=1}^{L-1}\tau_i^z \tau_{i+1}^z -\sum_{i=1}^{L} h_i \tau_i^x,
 \label{Hrand}
\end{equation}
where $\tau_{i}^{\mu}$ are the Pauli matrices at site $i$, and $h_i$ is a transverse magnetic field that has now become site dependent, according to the choice made for $(s_i^x)^2$. Specifically, one has $h_i=h\, |s_i^x|$. 

This means that if one randomly selects the values of the conserved quantities for the whole ladder (out of the available possibilities), the resulting Hamiltonian will be an Ising one with uniform coupling constant $J$ and with a transverse magnetic field that in general looks random. The values of $h_i$ are distributed independently on each site, and the probabilities follow from the multiplicities of the different values
\begin{equation}
P[h_i=h\, (M-2k )]=\frac{\nu_{M-2k}}{\sum_{k=0}^{\floor*{\frac{M}{2}}}\nu_{M-2k}}\, .
\label{probability}
\end{equation}
For an explicit example we refer to Appendix \ref{examplemultiplicity}.

\begin{figure}
\centering
\includegraphics[scale=0.29]{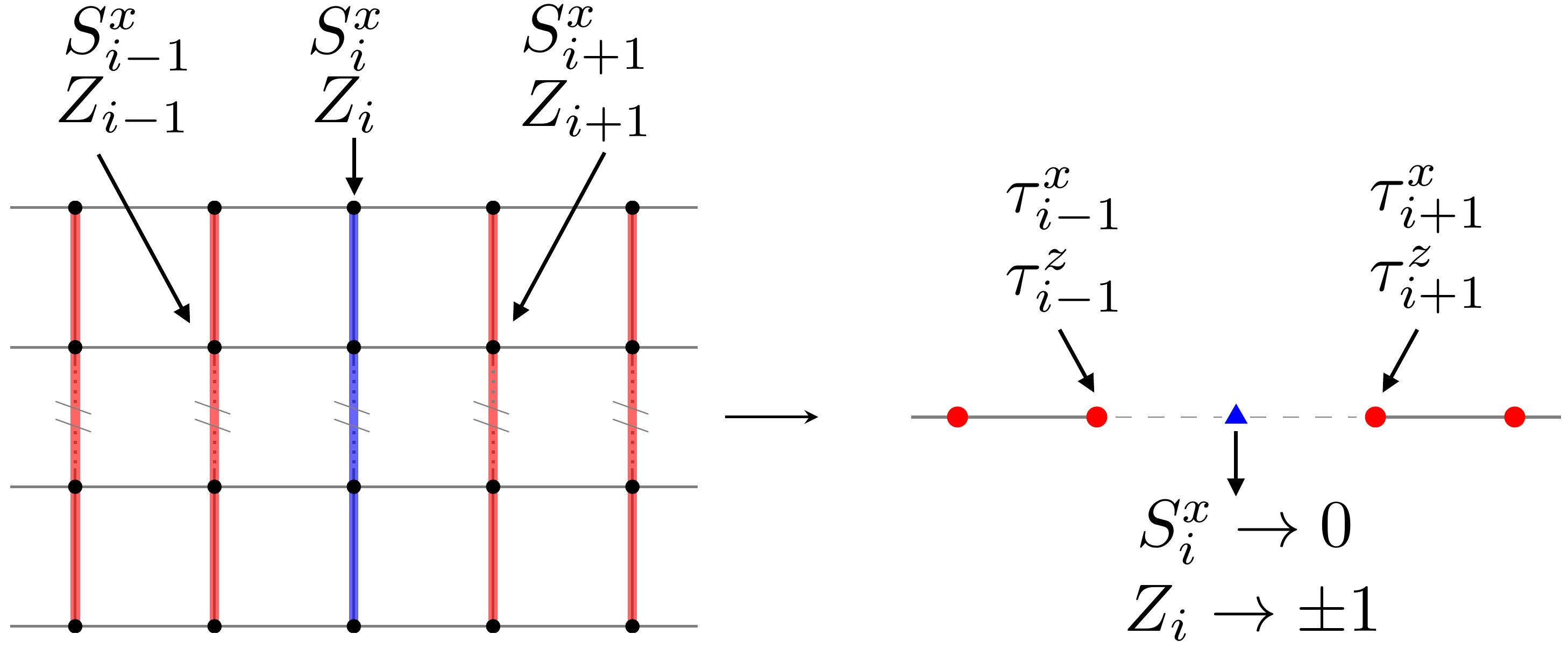}
\caption{{\bf Mapping of the ladder model for even $M$ with $(S_i^x)^2=0$ in one rung}.
  The representative operators, appearing in the Hamiltonian, for each rung are $Z_i$ and $S_i^x$; when $(S_i^x)^2\neq 0$ (red color) these operators are mapped into operators proportional to Pauli matrices $\tau_i^z$ and $\tau_i^x$. 
  On the other hand, when $(S_i^x)^2=0$ (blue color) then one has the mapping $S_i^x\to 0$ and $Z_i\to \pm 1$. The Hamiltonian in this sector is simply recovered by substituting the operators $Z_i$ and $S_i^x$ with the appropriate quantities, see Eq.~\eqref{Hcut}. }
\label{ladder_exp}
\end{figure}

\subsection{Block decomposition of the ladder Hamiltonian for $M$ even}
 
When the number of spins $M$ on each rung is even, the situation is slightly more complicated than the one for $M$ odd. While also for this case, fixing for each rung the values  $(s_i^2, (s_i^x)^2)$, the ladder Hamiltonian is formally mapped onto an Ising Hamiltonian [{\it cf}. Eq.~\eqref{Hrand}], unlike the $M$ odd case, now the site dependent transverse magnetic field $h_i$ can assume also the value $0$, when $s_i^x=0$.
Moreover, also $Z_i$ needs to be treated with some extra care: this operator can be mapped onto a Pauli matrix at site $i$, $\tau_{i}^{z}$, when $s_i^x \ne 0$, whereas it is mapped into $\pm1$ when $s_i^x=0$; the sign must be consistent with the symmetric or antisymmetric property of the state of the rung with zero magnetization along the $x$ direction. Apart from this, the probability of obtaining a given value of $(s_i^x)^2$ when randomly choosing a sector for the $i$th rung is still described by Eq.~\eqref{probability}. 

To better understand how the mapping works in this case we now discuss a simple example: let us consider the case where the value $(s_i^x)^2$ is zero only at one site, namely $s_i^x=0$, and  $s_j^x \ne 0 \,\, \forall j\ne i$. The ladder Hamiltonian is thus mapped onto the following Hamiltonian, see Fig.~\ref{ladder_exp},
\begin{align}
  \tilde{H}= & -J \sum_{j=1}^{i-2}\tau_j^z \tau_{j+1}^z-J \sum_{j=i+1}^{L-1}\tau_j^z \tau_{j+1}^z -  \sum_{j=1}^{i-1} h_j \tau_j^x 
  \nonumber \\
  &- \sum_{j=i+1}^{L} h_j \tau_j^x \pm\left(  J \tau^z_{i-1} + J \tau^z_{i+1}\right) .
  \label{Hcut}
\end{align}
It is clear here that the transverse field for the site $i$ is turned off, and the sign in front of the last term in the above equation is determined by the symmetry property of the state associated to the zero eigenvalue of $S_i^x$: an antisymmetric state will result in a minus sign while a symmetric one will lead to a plus. Eq.~\eqref{Hcut} also shows that, in this sector, the ladder system is broken up into two separate chains that are disconnected by the $i$-th site. 

 Notice that, contrary to the odd case where all sectors have the same dimension $2^L$ (corresponding  indeed to the dimension of a TFIM) in the $M$ even case, the dimension of each sector varies. This can be  understood by looking at relations \eqref{rungmapping}: whenever one has $(s_i^x)^2=0$ on a rung then the dimension of the sector is reduced of a factor $2$. As an example in a sector with exactly $n$ rungs assuming a zero value of the magnetization along the $x$ direction, the dimension is $2^{L-n}$.

\section{Long coherence times of boundary spins}

In the previous section we have shown how the ladder Hamiltonian assumes a block diagonal form and how, in each of these blocks, it can be represented as an Ising-like Hamiltonian
with values of the transverse field that depend on the conserved quantum numbers identifying the chosen sector. 
In addition, we have characterized the probabilities associated with these different values of the transverse field in the case in which a set of conserved quantities for each rung of the chain is randomly picked; 
practically speaking a  random selection of a set of these conserved quantum numbers can mimic the presence of disorder in the transverse field profile of an open boundary Ising model, with probabilities for the single site fields $h_i$ given by relation \eqref{probability}.

Even in the case of inhomogeneous transverse field, as for ordered Ising models, it is possible to construct a SZM operator $\tilde \Psi$; one way to proceed is to apply the usual iterative procedure \cite{Kemp, Fendley, Fendley1}. 
On the other hand it is also possible to obtain the representation of SZM for the chosen set of conserved quantities directly from Eq.~\eqref{SZM} exploiting the same relations in Eq.~\eqref{rungmapping} used to obtain the Hamiltonian.

Collecting all fixed choices for the quantum numbers $(s_i^2,(s_i^x)^2)$ into a vector $\vec{\omega}$ which now uniquely identifies the symmetry sector of the Hamiltonian, we can write the SZM $\Psi_{\vec{\omega}}$ in the sector as 
\begin{equation}
 {\Psi}_{\vec{\omega}}= \sum_{j=1}^L  \tau_j^z \prod_{k=1}^{j-1} \frac{h_k}{J } \tau_{k}^x   \, . 
\label{SZMsector}
\end{equation}
We can now make two interesting observations: first, we notice that the condition for the existence of this SZM is less restrictive than the condition implied by Eq.~\eqref{norm} for the existence of the SZM in the ladder. 
Indeed, in this sector one just needs
$$
\mathcal{N}_{\vec{\omega}}^2=\sum_{j=1}^\infty \prod_{k=1}^{j-1}\left(\frac{h_k}{J}\right)^2<\infty
$$
which can be true even if $\alpha=Mh/J\ge1$. One can thus have sectors where a SZM can be defined even if the full system cannot support one. This is illustrated in Fig.~\ref{Fig2}: the region labelled I corresponds to $\alpha<1$ where the overall system has a SZM; the region labelled II corresponds to $\alpha>1$ but where some sectors have normalisable SZM operators \eqref{SZMsector}; region III is where no sector has SZMs. 

The second observation concerns only $M$ even. In this case, the value of $(s_i^x)^2$ can be zero for some rungs of the ladder and this results in the presence of an {\it exact} SZM, \ie, an operator which strictly commutes with the Hamiltonian. 
For instance, if the the first rung of the ladder with $(s_i^x)^2=0$ is the $m$-th one, then, since $S_m^x\to 0$ in this sector, the exact SZM is still given by relation \eqref{SZMsector}, but with the sum over the rungs that  runs only up to the $m$-th term. In addition, looking at the term $B_L$ of equation \eqref{TelSum} the substitution $S_m^x\to 0$ also indicates that the SZM must exactly commute with the Hamiltonian, see Appendix \ref{compactform}.

\begin{figure}
\centering
\includegraphics[scale=1]{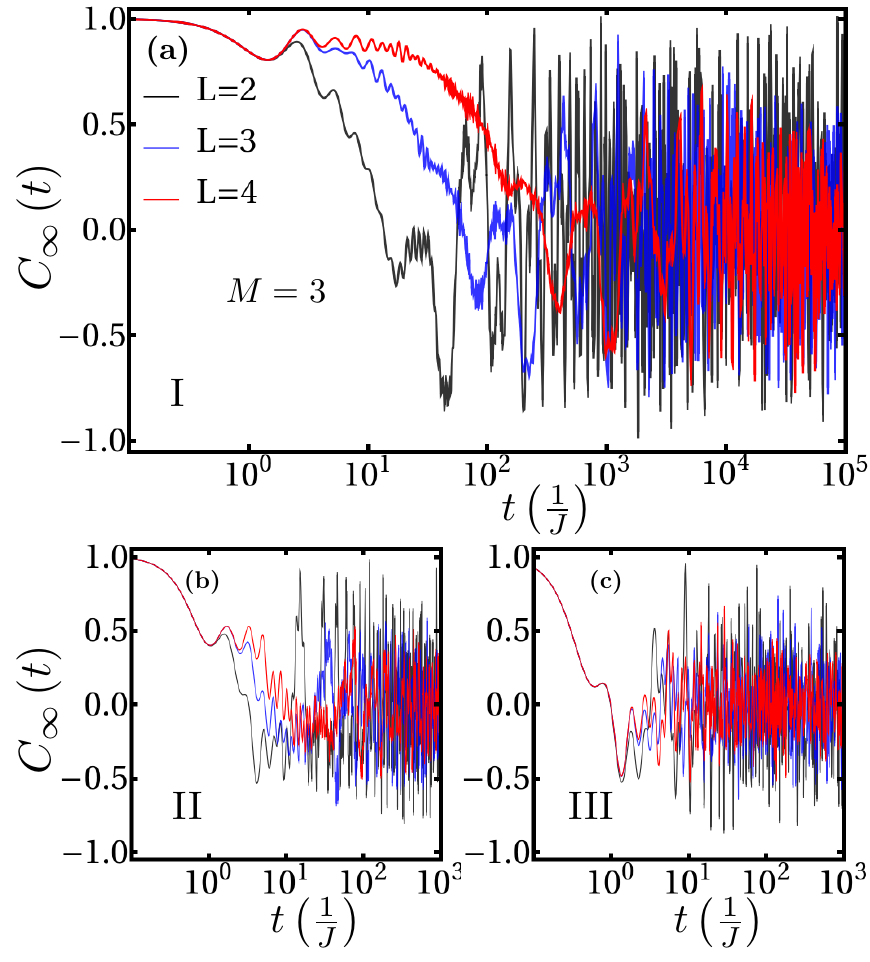}
\caption{ {\bf Infinite temperature edge autocorrelation from exact diagonalization for $M$ odd}. 
 {\bf (a) } Here we have taken $J=1$ and $h=0.2$ and compared the behavior of the infinite temperature time correlation functions for the edge spin operator $Z_1$ [\cf Eq.~\eqref{ACF}] in the ladder model with $M=3$ for different system sizes. Increasing the length $L$ of the ladder we see how the correlation time increases exponentially. 
At later times we expect these to approach a zero value with chaotic oscillations that should be damped for larger and larger system sizes.
 {\bf (b)} For $J=1$, $h=0.5$ and same $L$ as in {\bf (a)} the infinite temperature time correlators decay at earlier times compared with those in {\bf (a)}
 since in this regime only some sectors of the ladder Hamiltonian possess a SZM [{\it cf}. Fig.~\ref{Fig2}].
 {\bf (c)} When $J=1$ and $h=1.1$ there is no sector featuring the presence of a SZM and this results in an almost immediate decay of the correlation functions (data shown for the same values of $L$ as in {\bf(a)}). Sectors I, II, III, correspond to those of Fig.~\ref{Fig2}.}
\label{Fig4}
\end{figure}

\begin{figure}
\centering
\includegraphics[scale=0.5]{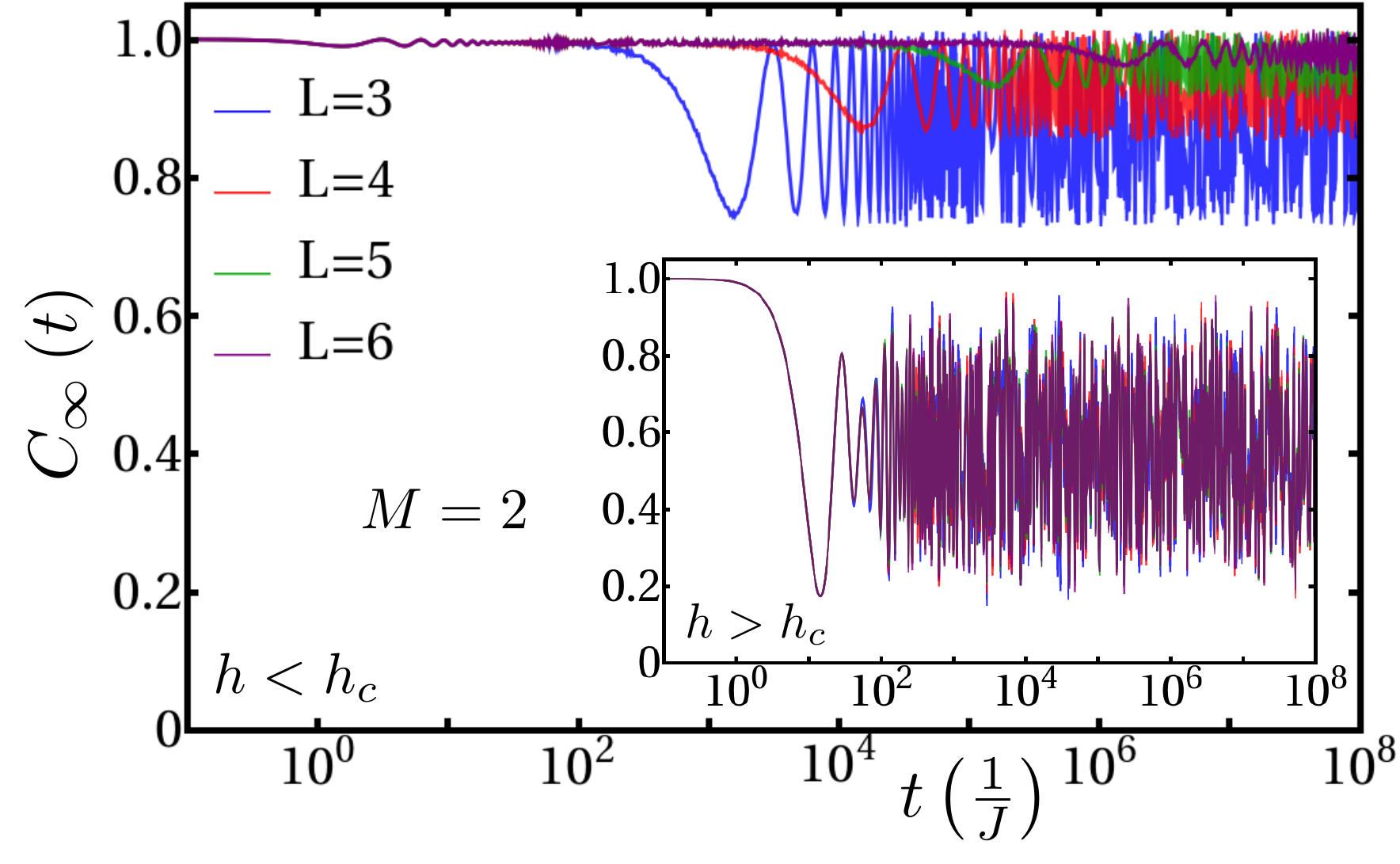}
\caption{{\bf Infinite temperature edge autocorrelation from exact diagonalization for $M$ even}. 
In this case we have considered $J=1$ and $h=0.05$ and compared the behavior  of the infinite temperature time correlation functions for the edge spin operator $Z_1$ [\cf Eq.~\eqref{ACF}] in the ladder model with $M=2$ for different system sizes. 
The value of the autocorrelator is not decaying to zero but actually saturates to a finite value. For the chosen parameters ($\alpha<1$), the remaining oscillations around this value decrease for increasing system sizes. 
In the inset we display the behavior of the infinite temperature time correlation functions when $J=0.05$, $h=0.05$ (resulting in $\alpha=2$) and $L$ as in the main figure. In this case, the autocorrelator saturates to a smaller value than in the previous case and the oscillations around this value are not affected by increasing the size of the system. }
\label{Fig5}
\end{figure}

These considerations also suggest the general behavior that should be displayed by the infinite temperature autocorrelation function \eqref{ACF}.
Indeed, we can think of the autocorrelator as given by the sum of contributions from all different sectors:
\begin{equation}
C_\infty(t)=\frac{1}{\eta}\sum_{\forall \vec{\omega}}\frac{1}{ \mbox{dim}_{\vec{\omega}}}\Tr\left(e^{it\tilde{H}_{\vec{\omega}}}\tau_1^{z}e^{-it\tilde{H}_{\vec{\omega}}}\tau_1^{z}\right)\, ,
\end{equation}
where $\eta=(\sum_{k=0}^{\floor*{\frac{M}{2}}}\nu_{M-2k})^L$
 is the total number of symmetry sectors in which the ladder Hamiltonian can be represented as an Ising-like Hamiltonian,  while $\mbox{dim}_{\vec{\omega}}$ is the dimension of the sector.

Exploiting a perturbative argument where one considers $\alpha\ll1$, similarly to what is done in \cite{JFL}, we can observe that in each sector the expected behavior of the autocorrelator is given by 
\begin{equation}
\frac{1}{\mbox{dim}_{\vec{\omega}}}\Tr\left(e^{it\tilde{H}_{\vec{\omega}}}\tau_1^{z}e^{-it\tilde{H}_{\vec{\omega}}}\tau_1^{z}\right)\approx \cos\left(t\, \Omega_{\vec{\omega}}\right) \, ,
\label{pert}
\end{equation}
where 
\begin{equation}
\Omega_{\vec{\omega}}=2 J \prod_{k=1}^L  \left( h_k / J \right)
\end{equation}
is a frequency (exponentially small in the system size) that depends on the profile of the transverse field. 
In the $M$ odd case, this frequency is never zero for finite systems, and the sum of all possible sectors can be understood as an average over the possible realizations of the transverse field profile. 
One therefore expects a flat behavior of the total autocorrelator up to time-scales of the order $\alpha^{-L}$ followed by a decay due to the average over all sectors. 
Indeed, after this time scale, one expects the oscillations due to the different frequencies of all the sectors to eventually become dephased. This trend is shown in Fig.~\ref{Fig4}(a): even if the considered system sizes are small, we see how the edge spins correlation times increase with the system size. After staying almost constant for times of order $\alpha^{-L}$, the autocorrelator, rather than displaying stable oscillations as in the TFIM, decays to zero. 

In Fig.~\ref{Fig4}(b) we display the same curves for a different value of the ladder transverse field $h$. 
In particular, we choose a value of $h$ for which there is no SZM for the whole ladder system but the operators \eqref{SZMsector} exist for some sectors, so that those sectors contribute long coherence times to the autocorrelator. Also in this case a dependence on the system size of the autocorrelator can be appreciated as a consequence of the fact that some of the sectors are featuring the presence of SZMs. 
On the other hand Fig.~\ref{Fig4}(c) shows that, for values of the ladder transverse field $h$ for which none of the sectors can sustain a SZM, the edge spins coherence time has no dependence on the system size and decay is fast. 

A different behavior is observed when $M$ is even. In this case, the fact that in many sectors $\Omega_{\vec{\omega}}=0$ prevents the autocorrelator to fully decay to zero and converging instead to a finite value. When $\alpha\ll 1$, the SZM has a large overlap with the edge operator $Z_1$ and the saturation value of the autocorrelator to be very close to $1$, as shown in Fig.~\ref{Fig5}. 
Furthermore, even when $\alpha>1$, because of the presence of exactly conserved SZMs in several sectors, the autocorrelator does not decay to zero either, but in contrast to $\alpha <1$ the decay to the asymptotic value is fast and size independent. Compare also the amplitude of the oscillations around the asymptotic value which seem to decrease with size for $\alpha <1$, \cf Fig.~\ref{Fig5} and its Inset.

\begin{figure}
\centering
\includegraphics[scale=1]{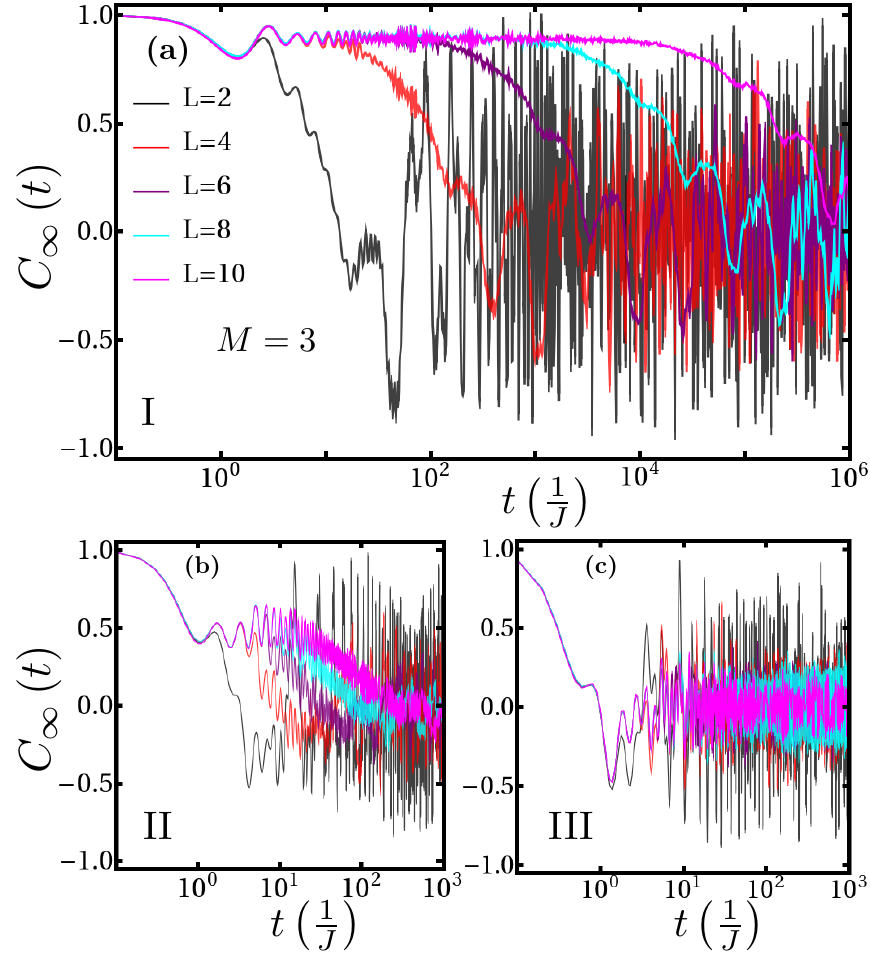}
\caption{ {\bf Infinite temperature edge autocorrelation from sampling from sector sampling for $M=3$. }
{\bf (a) }Also in this case we have taken $J=1$ and $h=0.2$ ($h_i=h |s_i^x|$) and compared the behavior of the infinite temperature time correlation functions for the edge spin operator $\tau^z_1$  [\cf Eq.~\eqref{rungmapping}] for different system sizes. 
For this choice of the parameters, where all the Ising models with random magnetic field over which the sampling is carried out have a SZM, 
by increasing system size it becomes more and more apparent the formation of a staircase like function.
{\bf (b)} For $J=1$, $h=0.5$ and same $L$ as in {\bf (a)}, a logarithmic decay of the infinite temperature time correlators can be observed which might be due to the fact that the above staircase like function shrinks.
{\bf (c)}  When $J=1$, $h=1.1$ and $L$ as in {\bf (a)} none of the Ising models with random magnetic field has SZM and the correlation functions which decay almost immediately do not display any size dependence. Sectors I, II, III, correspond to those of Fig.~\ref{Fig2}.}
\label{Fig6}
\end{figure}

The mapping of Eq.~\eqref{rungmapping} can be used to investigate the features of the correlation functions for larger system sizes than 
those shown in Figs.~\ref{Fig4} and \ref{Fig5} by exact diagonalization. 
Figure~\ref{Fig6} shows the correlation functions of the boundary spin operator $\tau^z_1$ \{\cf Eq.~(17) in \cite{Kemp}\} from sampling Ising models with random transverse fields compatible with the distribution of $M=3$, \cf \eqref{probability}. 
In Fig.~\ref{Fig6}(a) one can observe - in addition to the features already discussed in Fig.~\ref{Fig4} - what appears to be plateaus due to the distinct lifetimes of the SZMs in the various symmetry sectors.
Figure~\ref{Fig6}(b) shows that for a larger value of $h$ ($h_i=h |s_i^x|$), the infinite temperature time correlation functions still have a system size dependence. In this case rather than via plateaus they appear to decay logarithmically: this would correspond to the fact that in this regime not all sectors have SZMs \eqref{SZMsector} and their lifetimes are shorter than those in the regime of Fig.~\ref{Fig6}(a).
Lastly, Fig.~\ref{Fig6}(c) shows the autocorrelation in a regime similar to that of Fig.~\ref{Fig4}(c). Here none of the sectors have SZMs and decay to zero is fast and size independent, as expected.

\section{Conclusions}

We have studied strong zero modes in a class of spin ladders with plaquette interactions. 
Like in other models, the presence of SZM implies long coherence times for boundary degrees of freedom, and thus provide a general mechanism for long timescales in quantum many-body systems. 
Our results here are an addition to the growing list of findings in this area \cite{Kemp,Fendley,Fendley1,Alicea,Else,McGinley,Ippoliti,Carmele,JFL}. 
We were able to obtain the explicit form of the SZM operators - and understand the behaviour of edge correlators - because the models we consider, 
while appearing complex superficially, can be brought to simple and (almost) solvable forms by exploiting the large number of symmetries they possess. An interesting direction for future study would be to explore the existence of SZMs or related slow operators in more complex quantum plaquette models, in particular those closer to systems with fractons \cite{Nandkishore2018}.

\acknowledgments

The research leading to these results has received funding from a VC Scholarship for Research Excellence (LMV) and a Nottingham Research Fellowship (MM) from the University of Nottingham, and EPSRC Grants No.\ EP/N03404X/1 (FC and JPG) and  
EP/R04421X/1 (JPG).

\appendix

\section{Pairing due to the SZM}
\label{pairing}
In this appendix we provide some examples regarding the pairing of the spectrum due to the existence of SZMs.
For $M=1$ we recover the usual one dimensional TFIM, where the SZM $\Psi$ anticommutes with the parity operator $\mathcal{F}$ \eqref{eqsymmetry} and it
maps the spectrum in the odd sector to that in the even sector up to corrections exponentially small in system size \cite{Kemp}. 
If $\ket{\varphi_{+}}$ is an eigenvector of the Hamiltonian associated with the eigenvalue $\lambda_+$ belonging to the parity sector $\mathbb{P}_{+}$ then one can show that $\Psi|\varphi_+\rangle$ is an eigenvector of the Hamiltonian associated with an eigenvalue $\lambda_-$, exponentially close in system size to $\lambda_+$, belonging to the parity sector $\mathbb{P}_{-}$ 
\begin{equation*}
H \ket{\varphi_{+}}=\lambda_+ \ket{\varphi_{+}}, \quad H \Psi \ket{\varphi_{+}}\approx \lambda_+\Psi \ket{\varphi_{+}},
\end{equation*}
\begin{equation*}
 \mathcal{F} \ket{\varphi_{+}}= \ket{\varphi_{+}},\quad  \mathcal{F}\Psi\ket{\varphi_{+}}= -\Psi\ket{\varphi_{+}}\, . 
\end{equation*}

For $M=2$ the Hamiltonian commutes with the single chain parity operators $\mathcal{F}_1$ and $\mathcal{F}_2$ 
meaning that we can organize its eigenstates in four sectors $\mathbb{P}_{++},\mathbb{P}_{+-},\mathbb{P}_{-+},\mathbb{P}_{--}$.
Given an eigenstate of the Hamiltonian $\ket{\varphi_{\beta_1,\beta_2}}$ belonging to a parity sector $\mathbb{P}_{\beta_1,\beta_2}$ with $\beta_i=\pm$, and corresponding to an eigenvalue $\lambda_{\beta_1,\beta_2}$, one can show that the SZM $\Psi$ is mapping this vector onto an eigenstate of the Hamiltonian belonging to the parity sector $\mathbb{P}_{-\beta_1,-\beta_2}$. Indeed, exploiting the anticommutation relations of the SZM with $\mathcal{F}_i$ we have that 
\begin{equation}
\begin{split}
&\mathcal{F}_1\Psi\ket{\varphi_{\beta_1,\beta_2}}=-\beta_1\Psi\ket{\varphi_{\beta_1,\beta_2}}\, , \\
&\mathcal{F}_2\Psi\ket{\varphi_{\beta_1,\beta_2}}=-\beta_2\Psi\ket{\varphi_{\beta_1,\beta_2}}\, , 
\end{split}
\end{equation}
and because of the almost commutation relation of the SZM with the Hamiltonian we have that $H\Psi\ket{\varphi_{\beta_1,\beta_2}}\approx \lambda_{\beta_1,\beta_2}\Psi\ket{\varphi_{\beta_1,\beta_2}}$. The latter relation shows that $\lambda_{\beta_1,\beta_2}$ is exponentially close to an eigenvalue in the parity sector $\mathbb{P}_{-\beta_1,-\beta_2}$. 

When $M>2$ the number of parity sectors increases as $2^M$ but with the same steps illustrated above one can show that the SZM pairs the spectrum of $H$ in the sector $\mathbb{P}_{\vec{\beta}}$ with that in $\mathbb{P}_{-\vec{\beta}}$.

\section{Derivation of the SZM for the ladder system}
\label{compactform}

In this appendix we first derive the compact expression of the SZM for the ladder model following the iterative procedure used in \cite{Kemp, Fendley} 
and then we show that the correction term to the commutator of the SZM with the system Hamiltonian is exponentially small in system size.\\
In order to obtain a compact expression for the SZM we start by noticing that the Hamiltonian can be split in two parts
\begin{equation}
 H_0=-J\sum_{i=1}^{L-1} \prod_{k=1}^{M}  \sigma_{i,\,k}^z \, \sigma_{i+1, \,k}^z \, ,
\end{equation}
and 
\begin{equation}
 H_1= -h \sum_{i=1}^L\sum_{k=1}^{M}  \sigma_{i,\,k}^x \, ,
\end{equation}
such that $ H=H_0+H_1 $.
We set the zeroth order contribution to the SZM to be $\Psi_1= \prod_{k=1}^{M}\sigma_{1,k}^z$ since it commutes with $H_0$, $\left[ H_0,\Psi_1 \right]=0$.
$\Psi_1$ does not commute with $H_1$ though
\begin{equation*}
\begin{split}
 \left[ H_1, \Psi_1 \right]&=2ih \bigg( \sigma^y_{1,1} \prod_{k=2}^{M}\sigma_{1,k}^z
 + \sigma^y_{1,2}\prod_{k=1,k\ne2}^{M}\sigma_{1,k}^z \\
 &+\sigma^y_{1,3}\prod_{k=1,k\ne3}^{M}\sigma_{1,k}^z\bigg),
 \end{split}
\end{equation*}
therefore we need to find a correction of order $h$ to the SZM that cancels $\left[ H_1, \Psi_1 \right]$ when commuted with $H_0$;
$\Psi_2=\left( h/J \right) \left( \sum_{k=1}^M \sigma^x_{1,k} \right) \prod_{k=1}^M \sigma^z_{2,k}$ works fine since
\begin{equation*}
 \left[ H_0, \Psi_2 \right]= - \left[ H_1, \Psi_1 \right].
\end{equation*} 
Now the commutator of $\Psi_2$ with $H_1$ will generate a contribution of order $h^2$ which can be canceled by a third correction to the SZM of order $h^2$.
Following this pattern it is easy to show that if the SZM has the following compact form
\begin{equation}
 \Psi= \sum_{j=1}^L \left(\frac{h}{J} \right)^{j-1} \prod_{i=1}^{j-1} \left( \sum_{k=1}^M \sigma_{i,k}^x \right) \prod_{k=1}^{M}\sigma_{j,k}^z \, ,
\label{SZMprova}
\end{equation}
then the norm of the error term to the commutator decays exponentially with system size, namely $ \| \left[H, \Psi \right]\| \approx \alpha ^L$ where $\alpha=M h /J$.\\
In order to prove this last statement and the form of Eq.~\eqref{SZMprova} we write the commutator, as mentioned in the main text, as a telescoping series 
\begin{equation}
  \left[ H, \Psi \right] = \sum_{j=1}^{L-1} \left(  A_{j+1} + B_j \right) + B_L \, ,
\end{equation}
and we now show that $A_{j+1}=- B_j, \forall j$, meaning that the only contribution to the commutator comes from the term $B_L$. We start by evaluating the term $B_j= \left[ H_1, \Psi_j \right]$.
% \begin{equation*}
%  \begin{align}
%   \left[ H_1, \Psi_j \right]= \bigg[-h \sum_{i=1}^L\sum_{k=1}^{M}  \sigma_{i,\,k}^x, 
%  \left(\frac{h}{J} \right)^{j-1} \prod_{l=1}^{j-1} \left( \sum_{n=1}^M \sigma_{l,n}^x \right) \prod_{n=1}^{M}\sigma_{j,n}^z \bigg].
%  \end{align}
% \end{equation*}
Using the fact that Pauli matrices on different sites act on different spin states we can see that the only non zero contribute to this commutator is the one when $\sigma^x_{i,k}$ and $\sigma^z_{j,n}$ act on the same site, 
that is $i=j$

 \begin{align*}
  B_j= \left[ H_1, \Psi_j \right]&= -  J \left( \frac{h}{J} \right)^{j} \prod_{l=1}^{j-1} \left( \sum_{n=1}^M \sigma_{l,n}^x \right)\\
   &\times \left[ \sum_{k=1}^{M}  \sigma_{j,\,k}^x \,,\prod_{n=1}^{M}\sigma_{j,n}^z \right].
 \end{align*}

On the other hand when we evaluate the term $A_{j+1}=\left[ H_0, \Psi_{j+1} \right]$,
% \begin{equation*}
% \left[ H_0, \Psi_{j+1} \right]= 
%   \left[ -J\sum_{i=1}^{L-1} \prod_{k=1}^{M}  \sigma_{i,\,k}^z \sigma_{i+1, \,k}^z, 
%   \left(\frac{h}{J} \right)^{j} \prod_{l=1}^{j} \left( \sum_{n=1}^M \sigma_{l,n}^x \right) \prod_{n=1}^{M}\sigma_{j+1,n}^z \right] \,,
% \end{equation*}
we observe that all the plaquette terms of $H_0$ commute with $ \Psi_{j+1} $ except the $j-$th plaquette
% \begin{equation*}
%  \begin{align}
%   \left[ H_0, \Psi_{j+1} \right]&= 
%   \left[ -J \prod_{k=1}^{M}  \sigma_{j,\,k}^z \,\sigma_{j+1, \,k}^z, \left(\frac{h}{J} \right)^{j} \prod_{l=1}^{j} \left( \sum_{n=1}^M \sigma_{l,n}^x \right) \prod_{n=1}^{M}\sigma_{j+1,n}^z \right] \\
%   &= -J \left(\frac{h}{J} \right)^{j} \left[  \prod_{k=1}^{M} \sigma_{j,\,k}^z ,\prod_{l=1}^{j} \sum_{n=1}^M \sigma_{l,n}^x \right] \\
%   &= -  J \left(\frac{h}{J} \right)^{j} \prod_{l=1}^{j-1} \left( \sum_{n=1}^M \sigma_{l,n}^x \right) \left[ \prod_{k=1}^{M} \sigma_{j,\,k}^z , \sum_{n=1}^M \sigma_{j,n}^x \right]
% \end{align}
% \end{equation*}

 \begin{align*}
 A_{j+1}= \left[ H_0, \Psi_{j+1} \right]&= 
   -J \left(\frac{h}{J} \right)^{j} \left[  \prod_{k=1}^{M} \sigma_{j,\,k}^z ,\prod_{l=1}^{j} \sum_{n=1}^M \sigma_{l,n}^x \right] \\
  &= -  J \left(\frac{h}{J} \right)^{j} \prod_{l=1}^{j-1} \left( \sum_{n=1}^M \sigma_{l,n}^x \right) \\
  &\times\left[ \prod_{k=1}^{M} \sigma_{j,\,k}^z , \sum_{n=1}^M \sigma_{j,n}^x \right]
\end{align*}

therefore 
\begin{equation*}
 A_{j+1}+B_j=0, \, \forall j.
\end{equation*}
To show that the norm of $B_L$ decays exponentially with system size we note that,
from all the terms of $H_1$, only the one at site $\left( j, k \right)$ gives non zero contribute
\begin{equation*}
 \left[ H_1, \Psi_j \right]= \left[-h \sum_{k=1}^M \sigma_{j, \, k}^x, \Psi_j \right]
\end{equation*}
and by using the Cauchy$-$Schwarz inequality $\| \left[ A,B \right]\| \le 2 \|A \| \|B \|$ and $ \| \Psi_j\| \le  \left( h/J \right)^{j-1} M^{j-1}$ we get
\begin{equation*}
\|B_L\|= \| \left[ H, \Psi \right]\| = \| \left[ H_1, \Psi_L \right]\| \approx 2   J \alpha ^L \,,
\end{equation*}
with $\alpha=M h/J$.

\section{Examples for the Clebsch$-$Gordan decomposition series}
\label{exampleseries}
In this appendix  we present two explicit examples, one for $M$ odd and one for $M$ even, for the Clebsch$-$Gordan decomposition series.\\
For $M=3$ spin-1/2 systems the series is

\begin{align*}
 \pmb{2}^{\otimes 3} &= \bigoplus \limits_{k=0}^{\floor*{1}} \left( \frac{4-2k}{4} \binom{4}{k} \right) \left( \pmb{4 -2k} \right) \\
 &= 1 \left( \pmb{4} \right)+ 2 \left(\pmb{2} \right),
\end{align*} 
which means that by composing three spins $1/2$ we will obtain one irreducible representation of dimension $4$ (i.e., $s_i^2= 4 \ell(\ell+1)$ where $\ell=3/2$ and  $s_i^x=-3,-1, 1, 3 $)
and one irreducible representation of dimension $2$ (i.e., $\ell= 1/2, s_i^x=-1, 1 $) with multiplicity $2$.
In terms of the quantum numbers $\left( s_i^2, (s_i^x)^2 \right)$ associated to the conserved quantities we will obtain $4$ doublets.

When composing $M=4$ spins $1/2$ the decomposition series gives
\begin{align*}
\pmb{2}^{\otimes 4} &= \bigoplus \limits_{k=0}^{\floor*{2}} \left( \frac{5-2k}{5} \binom{5}{k} \right) \left( \pmb{5-2k} \right) \\
 &= 1( \pmb{5})+3 \, (\pmb{3})+2 \, (\pmb{1}),
\end{align*} 
one irreducible representation of dimension $5$ (i.e., $s_i^2= 4 \ell(\ell+1)$ where  $l=2$ and $s_i^x=-4,-2,0,2,4$ ),
one irreducible representation of dimension $3$ (i.e., $\ell=1, s_i^x=-2,0,2$ ) with multiplicity 3 and
one irreducible representation of dimension $1$ (i.e., $l=0, s_i^x=0$) with multiplicity 2.
In terms of the quantum numbers $\left( s_i^2, (s_i^x)^2 \right)$ associated with the conserved quantities  we will obtain $5$ doublets and $6$ singlets.

\section{Probability for the random transverse field}
\label{examplemultiplicity}
In Appendix \ref{exampleseries} we provided the multiplicities and dimension of all irreducible representations one can obtain when combining M spin $1/2$ systems for both $M$ even and $M$ odd cases. 
Here we want to explicitly write all possible values of $|s_i^x|=M-2k$ and their multiplicities $\nu_{M-2k}$ [{\it cf}. Eq.~\eqref{multiplicity}] for $M=3$ and $M=4$.\\
When composing $M=3$ spins $1/2$ the values that $|s_i^x|=M-2k$ can assume are $|s_i^x|=3$ with multiplicity $\nu_{3}=1$ and  $|s_i^x|=1$ with multiplicity $\nu_{1}=3$.\\
For $M=4$ the values assumed by $|s_i^x|=M-2k$ with their multiplicities are $|s_i^x|=4$ with $\nu_{4}=1$,  $|s_i^x|=2$ with $\nu_{2}=4$ and $|s_i^x|=0$ with $\nu_{0}=6$.
This information allows one to calculate the probability for the random transverse magnetic field.
For instance for $M=3$ [{\it cf}. Eq.~\eqref{probability}]
\begin{equation*}
 P[h_i=3 \,h]= \frac{1}{4}, \quad P[h_i=h ]= \frac{3}{4},
\end{equation*}
there is probability $1/4$ to have the value $3h$ and probability $3/4$ to have the value $h$ while 
 for $M=4$
\begin{align*}
 P[h_i=4\,h ]&= \frac{1}{11}, \quad P[h_i=2\,h ]= \frac{4}{11}, \\
 P[h_i=0]&= \frac{6}{11},
\end{align*}
there is probability $1/11$ to have the value $4h$,  $4/11$ to have the value $2h$ and  $6/11$ to have the value $0$ for $h_i$.

%  \section{Symmetry properties of the Clebsch-Gordan coefficients}
%  \begin{equation*}
%  \braket{j_1 m_1 \,j_2 m_2| JM}= (-1)^{j_1 +j_2 -J} \braket{j_2 m_2\, j_1 m_1| JM}
%  \end{equation*}
%  where $j_1$ and $j_2$ represent the different angular momenta quantum numbers of two systems, 
%  while $m_1$ and $m_2$ are the angular momenta projection on the $x$ axis. 
%  $J$ and $M$ are respectively the total angular momentum quantum number and its $x$ component.

\bibliography{LadderSZM}
\bibliographystyle{apsrev4-1} %questo commando non mostra i titoli dei paper citati
\end{document}